\documentclass[doublecol,nocite]{epl2}
\newcommand{\cP}{\ensuremath{\mathcal{P}}}
\newcommand{\cT}{\ensuremath{\mathcal{T}}}

\newcommand{\cPT}{\ensuremath{\mathcal{PT}}}

\title{Complex phases in quantum mechanics}
\shorttitle{Complex quantum phases} 

\author{Carl M. Bender\inst{1} \and Daniel W. Hook\inst{1,2}
}
\shortauthor{Carl M. Bender \etal}

\institute{                    
\inst{1} Department of Physics, Washington University, St. Louis, Missouri 63130, USA \\
\inst{2} Centre for Complexity Science, Imperial College London, London, SW7 2AZ, UK 
}

\abstract{Hamilton's equations of motion are local differential equations and boundary conditions are required to determine the solution uniquely. Depending on the choice of boundary conditions, a
Hamiltonian may thereby describe several different physically observable phases, each exhibiting its own characteristic global symmetry.}

\begin{document}
\maketitle

\section{We live in a complex world}
The mathematical theory of complex variables was developed centuries before the physical theories of gravity, electricity and magnetism, fluid mechanics, and relativity, but complex numbers did not appear in the formulation of physical theories before the advent of quantum mechanics. Indeed, the fundamental physical phenomena of propagation, equilibrium, and diffusion are described by hyperbolic, elliptic, and parabolic equations, all of which are real. Physicists have long used powerful complex-variable techniques to perform theoretical calculations, but the world was not thought to be influenced by what might be happening off the real axis and in the complex plane.

The development of quantum mechanics changed everything because the Schr\"odinger equation 
$$i\psi_t(x,t)=-\psi_{xx}(x,t)+V(x)\psi(x,t)$$
depends explicitly on $i$, so the quantum amplitude $\psi(x,t)$ is complex and calculating the real probability density $|\psi(x,t)|^2$ requires complex conjugation. 
The number $i$ is ubiquitous in quantum theory: The uncertainty principle follows from the operator commutation relation $[{\hat x},{\hat p}]=i$; fermionic representations are complex; tunneling involves propagation in complex space; as Wigner noted, the time-reversal operator $\cT$ performs complex conjugation.

A mathematical description of a complex universe is fundamentally different from that of a real universe because complex calculus differs from real calculus. In the complex plane the number $z=x+iy$ is a single variable and $\frac{d}{dz}$ is an {\it ordinary} derivative. In the real plane $(x,y)$ is a pair of real numbers, and derivatives with respect to $x$ or $y$ are {\it partial} derivatives. If the complex derivative $f'(z)$ of the function $f(z)$ exists, then all higher derivatives of $f(z)$ also exist. However, if the partial derivative, say $f_y(x,y)$, of a real function $f(x,y)$ exists, derivatives with respect to $x$ and higher derivatives with respect to $y$ may not exist.

\begin{figure}[h!]
\center
\includegraphics[scale = 0.75]{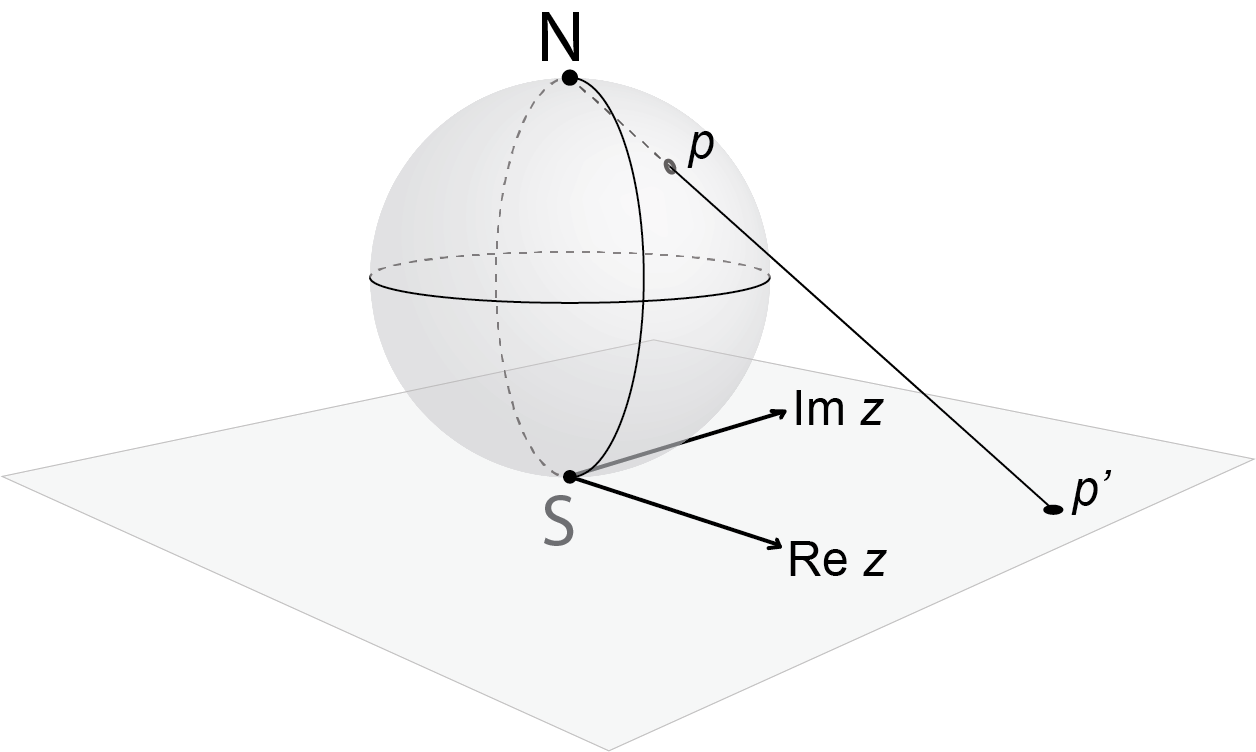}
\caption{1-to-1 mapping from a sphere to the complex plane.}
\label{Fig1}
\end{figure}

The topology of the complex plane is profoundly different from that of the real plane. The complex plane is {\it compact} and there is a unique point at complex $\infty$.
The compactness of the complex plane allows there to be a 1-to-1 {\it stereographic mapping} between each point $p$ on a sphere and a corresponding point $p'$ on the complex plane (see Fig.~\ref{Fig1}). The south pole of the sphere sits at $z=0$ in the complex plane. A straight line from the north pole on the sphere to a point on the plane passes through a unique associated point on the sphere. The north pole on the sphere corresponds to the unique point $z=\infty$.

Complex infinity is defined as $\infty\equiv\lim_{z\to0}\frac{1}{z}$, where the complex limit $z\to0$ may be taken along {\it any} path in the complex plane that approaches $z=0$. Because $z$ may approach 0 along the negative-real axis or the positive-real axis, it follows that negative-real infinity and positive-real infinity are the same point. In contrast, the real plane is not compact and $\infty$ is not a unique point; $(+\infty,0)$ on the $x$ axis is infinitely distant from $(-\infty,0)$.

\section{Complex classical mechanics}
An elementary classical-mechanical system illustrates the topological difference between the real plane and the complex plane. Consider the motion of a particle on the real-$x$ axis in an upside-down quartic potential. The nonrelativistic Hamiltonian for such a system is $H=p^2-x^4$ and Hamilton's equations of motion read \cite{pt486,pt579}
\begin{equation}\label{e1}
\dot x(t)=2p(t),\qquad\dot p(t)=4[x(t)]^3.
\end{equation}
If a particle is initially at the origin $x(0)=0$ and its initial energy is positive $E>0$, the time $T$ for this particle to slide down the potential hill and reach $x=\infty$ is {\it finite}:
\begin{equation}\label{e2}
T=\textstyle{\int_0^\infty} dx\, (4E+4x^4)^{-1/2}=0.927037...\,E^{-1/4}.
\end{equation}

This raises an interesting question: Where is the particle when $t>T$? There are two answers to this question. If the particle is strictly confined to the real line, it remains at $x=\infty$; it has nowhere else to go.

\begin{figure}[t]
\center
\includegraphics[scale = 0.25]{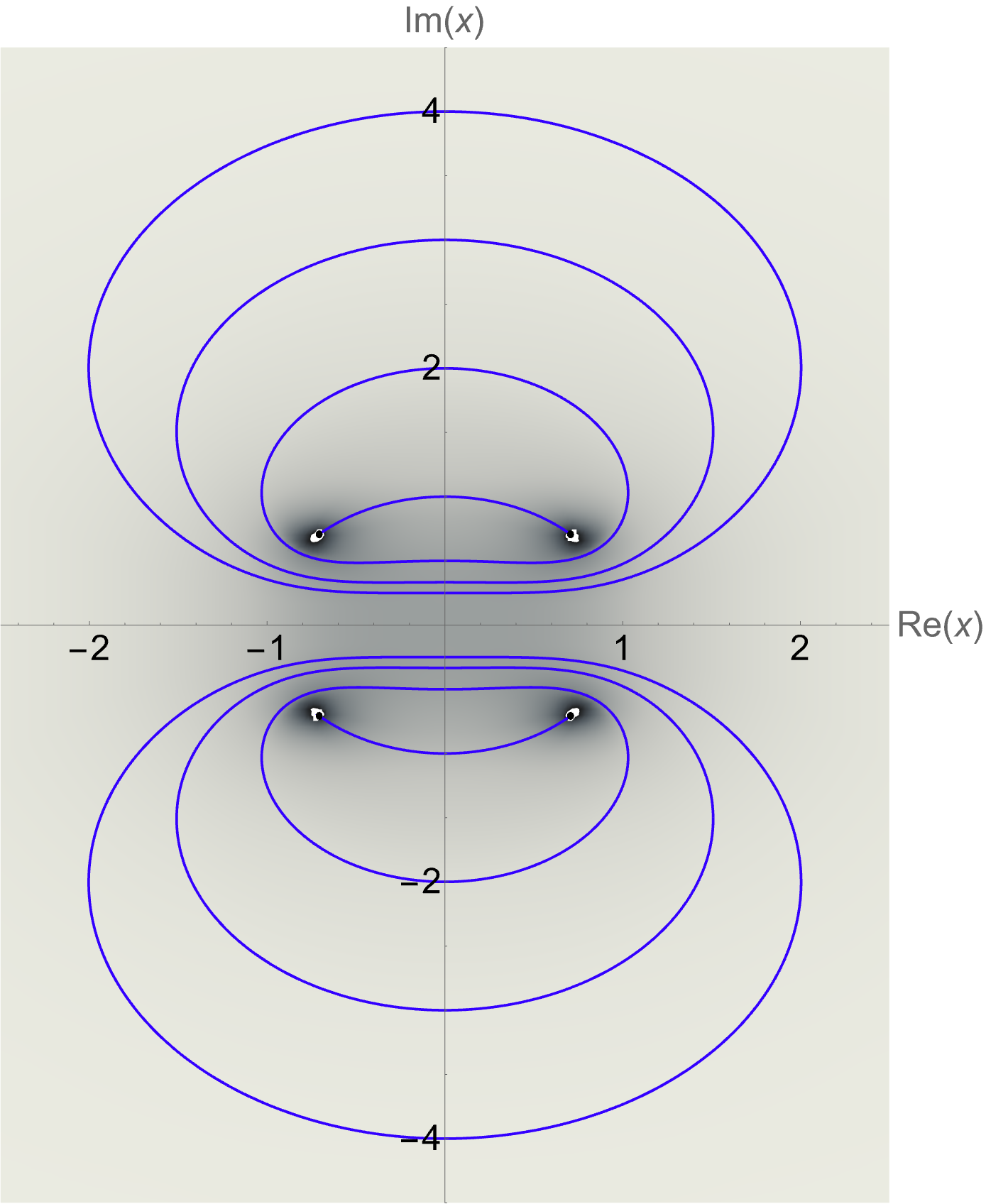}
\caption{Classical paths of a particle of energy $E=1$ subject to a $-x^4$ potential in the complex-$x$ plane. All orbits are closed and periodic with period $2T$ [see (\ref{e2})]. There are four turning points at a distance of 1 from the origin, two in the upper-half and two in the lower-half $x$ plane. All orbits are left-right symmetric ($\cPT$ symmetric). A particle at $x=0$ will run off to $\pm\infty$ and return to $x=0$ from $\mp\infty$ in time $2T$.}
\label{Fig2}
\end{figure}

However, if the real line is embedded in the {\it complex}-$x$ plane, we observe a totally different classical behavior. We treat $x(t)$ and $p(t)$ as {\it complex} functions of the parameter $t$. The classical equations of motion (\ref{e1}) determine the classical trajectories in the complex-$x$ plane (see Fig.~\ref{Fig2}). 

Figure~\ref{Fig2} shows two pairs of classical turning points, one in the upper-half and the other in the lower-half complex-$x$ plane. A classical particle may oscillate between each pair of turning points. The other orbits are periodic and enclose the pairs of turning points. The real axis is a {\it separatrix}; no trajectories may cross the real axis. Large orbits are roughly D-shaped and in the limit as the orbits increase in size, one side of the orbit approaches the real axis while the other makes a huge semicircular sweep from $\pm\infty$ to $\mp\infty$. Thus, a particle initially at the origin runs off to $\pm\infty$ and then returns to the origin from $\mp\infty$.

As a consequence of Cauchy's theorem and path independence, the period of every classical orbit in Fig.~\ref{Fig2} is $2T$. Thus, particles in larger orbits move faster. If a particle starts at the origin, it zooms off to $\pm\infty$ along the real-$x$ axis and it reaches infinity in time $T$. It then returns to the origin from $\mp\infty$, again in time $T$.

How did such a particle go from $\pm\infty$ to $\mp\infty$ in no time? The physical answer is that the particle travels at infinite speed at $|x|=\infty$. The mathematical answer is that the point at complex $\infty$ is unique because the complex plane is compact; so, if the particle is at $\pm\infty$, it is {\it already} at $\mp\infty$. The stereographic image of this path is quite simple: The particle begins at the south pole, travels on a great circle up to the north pole, and then continues along the great circle back to the south pole. This particle never changes its direction of travel; if it leaves the origin traveling to the right, it returns to the origin still traveling to the right.

The real-world and complex-world dynamical behaviors of a classical particle are profoundly different. A classical particle traveling along the $x$ axis in the {\it real} plane is in an {\it unbound} state; it goes from the origin to $\pm\infty$ and never comes back. This classical behavior resembles that of an unstable decaying quantum system located at $x=0$:
\begin{equation}
-\infty\,\longleftarrow\,0\,\longrightarrow\,\infty.
\label{e3}
\end{equation}

However, in the complex plane all orbits are periodic and return to their initial positions:
\begin{equation}
-\infty\,\longrightarrow\,0\,\longrightarrow\,\infty~~~{\rm and}~~~-\infty\,\longleftarrow\,0\,\longleftarrow\,\infty.
\label{e4}
\end{equation}
This system is {\it dynamically stable}. Because the particle is statistically most likely to be found where it is traveling slowly, the complex motion resembles that of a {\it localized} quantum {\it bound} state. The classical probability of finding the particle near $x=0$ in the complex plane is peaked at $x=0$ and vanishes like $x^{-2}$ for large $|x|$ (see \cite{pt579}).

Particle trajectories corresponding to these real-unstable and complex-stable behaviors exhibit different {\it global symmetries}. In (\ref{e3}) particles on the negative-real line are left-moving and particles on the positive-real line are right-moving. These trajectories are {\bf parity} $\cP$ symmetric; under $\cP$ reflection ($x\to-x$) right-moving particles for $x>0$ become left-moving particles for $x<0$.

However, in the complex plane every trajectory is {\bf parity-time ($\cPT$) symmetric}. This is because space reflection $\cP$ replaces $x+iy$ with $-x-iy$ and time reversal $\cT$ performs complex conjugation and thus it replaces $-x-iy$ with $-x+iy$. Combined $\cP$ and $\cT$ changes the sign of $x$, so $\cPT$ performs a horizontal (left-right) reflection about the imaginary axis. (The direction of motion in each orbit is also reversed under this $\cPT$ reflection because time reversal changes the sign of the velocity.)

While the $\cPT$ operator performs a reflection about the imaginary axis, Hermitian conjugation (complex conjugation) is a reflection about the real axis. Thus, these are two orthogonal and independent symmetry operations. In the complex plane these symmetry operations are distinct from parity $\cP$, which performs a reflection through the origin ($x+iy\to-x-iy$).

\section{Classical phases}
The example above shows that one Hamiltonian may generate completely different kinds of classical-mechanical behaviors, which we refer to as {\it phases}. In one phase particles on the real axis escape to $\pm\infty$, as illustrated schematically in (\ref{e3}). This phase is characterized by having global $\cP$ symmetry. Classical trajectories in this phase are invariant under parity reflection.

In the second phase particles follow periodic orbits. On the real axis these particles run off to $\pm\infty$ and come back from $\mp\infty$, as illustrated schematically in (\ref{e4}). This phase is globally $\cPT$ symmetric. All trajectories in this phase are parity-time symmetric and particle trajectories on the real axis in this phase are said to be {\it unidirectional} \cite{pt189}.

\section{Quantum phases} Quantum-mechanical Hamiltonians can also define multiple phases that are distinguished by having different global symmetries. For example, the quantum anharmonic-oscillator Hamiltonian with an upside-down quartic potential,
\begin{equation}
{\hat H}={\hat p}^2+{\hat x}^2-{\hat x}^4,
\label{e5}
\end{equation}
defines two phases. The time-independent Schr\"odinger equation associated with $\hat H$ in (\ref{e5}) is
\begin{equation}
-\psi''(x)+x^2\psi(x)-x^4\psi(x)=E\psi(x).
\label{e6}
\end{equation}

Physically observable solutions to (\ref{e6}) belong to two distinctly different phases because they obey different boundary conditions at $|x|=\infty$. Depending on how the negative sign in front of the quartic term in $\hat H$ is obtained, we get either a $\cP$-symmetric phase with complex-energy eigenvalues and unstable states, or a $\cPT$-symmetric phase for which the energy spectrum is entirely real and positive, and the states are all stable.

To obtain the $\cP$-symmetric phase of (\ref{e5}), we begin with the Hamiltonian for a quantum anharmonic oscillator with a {\it right-side-up} potential
\begin{equation}
{\hat H}={\hat p}^2+{\hat x}^2+g{\hat x}^4\quad(g>0).
\label{e7}
\end{equation}
The Hamiltonian (\ref{e7}) is parity symmetric; it is invariant under ${\hat x}\to-{\hat x}$ and ${\hat p}\to-{\hat p}$. We construct the upside-down quartic Hamiltonian in (\ref{e5}) from $\hat H$ in (\ref{e7}) by performing an analytic rotation in complex-coupling-constant space; that is, we let $g=e^{i\theta}$ and rotate smoothly in the positive sense from $\theta=0$ to $\theta=\pi$. Under this rotation the parity symmetry of $\hat H$ in (\ref{e7}) is preserved.

We must perform this $\theta$ rotation carefully: As we increase $\theta$, we must {\it simultaneously rotate the boundary conditions on the eigenfunctions} of $\hat H$ in (\ref{e7}). We cannot solve the Schr\"odinger equation (\ref{e6}) exactly, but WKB theory provides an asymptotic approximation to $\psi(x)$ for large $|x|$, and thus WKB specifies the boundary conditions that the eigenfunctions obey as the $\theta$ rotation is performed \cite{pt14}.

When $\theta=0$, the usual boundary conditions on the eigenfunctions $\psi(x)$ are simply that $\psi(x)\to0$ as $x\to\pm\infty$. However, the WKB approximation to $\psi(x)$ specifies that the eigenfunctions vanish exponentially for large $|x|$ like $\exp(-\frac{1}{3}|x|^3)$. These eigenfunctions also vanish exponentially as $|x|\to\infty$ in the complex-$x$ plane in a pair of wedge-shaped regions called {\it Stokes sectors}. These sectors are centered about the positive-real and negative-real axes and have an angular opening of $\frac{1}{3}\pi$. As $\theta$ increases from $0$ to $\pi$, this pair of wedges rotates clockwise like a propeller by an angle of $\frac{1}{6}\pi$ until at $\theta=\pi$ the upper edge of the right sector lies on the positive-$x$ axis and the lower edge of the left sector lies on the negative-$x$ axis.

We can also rotate $\theta$ in the negative sense from $0$ to $-\pi$. In this case the pair of Stokes sectors rotate anticlockwise by an angle of $\frac{1}{6}\pi$ until the lower edge of the right sector lies on the positive-$x$ axis and the upper edge of the left sector lies on the negative-$x$ axis. Both the positive and negative $\theta$ rotations preserve the $\cP$ symmetry of the original Hamiltonian (\ref{e7}); during the rotation process the pair of Stokes sectors remains symmetric under a reflection through the origin in the complex-$x$ plane.

\begin{figure}[t]
\center
\includegraphics[scale = 0.25]{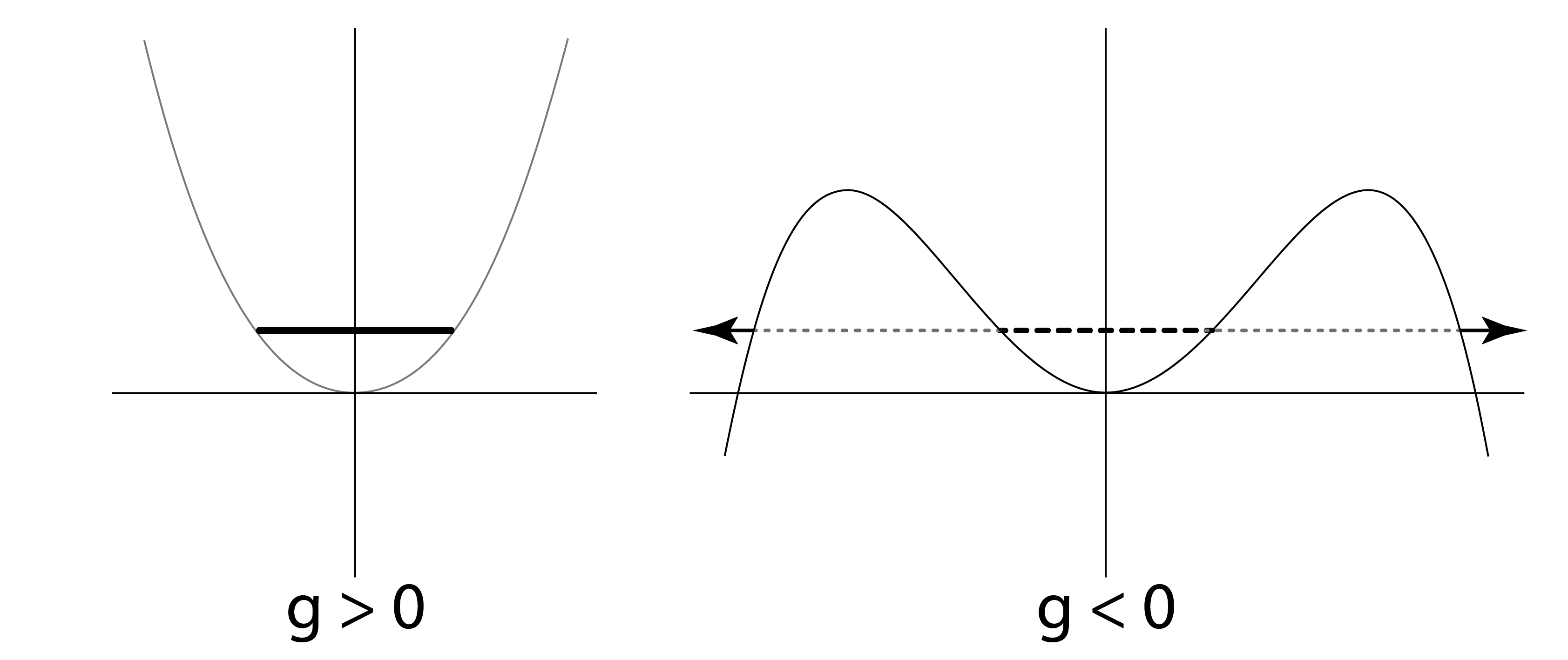}
\caption{Schematic diagram showing how a stable bound state in the right-side-up quartic potential ${\hat x}^2+e^{i\theta}{\hat x}^4$ at $\theta=0$ (left panel) becomes unstable as $\theta$ is rotated from $0$ to $\pi$. At $\theta=\pi$ waves propagate outward to $x=\pm\infty$ in a parity-symmetric fashion (right panel).}
\label{Fig3}
\end{figure}

For both positive and negative rotations of $\theta$, the edges of the Stokes sectors lie on the real-$x$ axis. This implies that the eigenfunction solutions to the Schr\"odinger equation (\ref{e6}) become wavelike (oscillatory). Depending on the choice of rotation direction, there are either outgoing-wave or incoming-wave solutions to the time-dependent Schr\"odinger equation. Figure \ref{Fig3} shows that a stable bound state in the right-side-up quartic potential (\ref{e7}) becomes unstable in the upside-down potential (\ref{e5}) and that the state tunnels outward to $\pm\infty$. This is the quantum analog of the classical behavior represented in (\ref{e3}). 

We can obtain the $\cPT$-symmetric phase of the upside-down potential in (\ref{e5}) by performing a smooth complex deformation of the harmonic-oscillator Hamiltonian. To do so, we use the complex non-Hermitian Hamiltonian
\begin{equation}
{\hat H}_\epsilon={\hat p}^2+{\hat x}^2+{\hat x}^2(i{\hat x})^\varepsilon\quad(\varepsilon~{\rm real}).
\label{e8}
\end{equation}
This Hamiltonian is $\cPT$ symmetric because a $\cP$ reflection changes the signs of the $\hat x$ and $\hat p$ and a $\cT$ reflection changes the signs of $\hat p$ and $i$. When $\varepsilon=0$, ${\hat H}_\varepsilon$ is the harmonic-oscillator Hamiltonian, which is both $\cP$ and $\cPT$ symmetric. As we smoothly increase $\varepsilon$ from 0 to 2, ${\hat H}_\varepsilon$ deforms into the Hamiltonian (\ref{e5}).

This smooth deformation breaks the $\cP$ symmetry but preserves the $\cPT$ symmetry of the harmonic-oscillator Hamiltonian. The harmonic-oscillator eigenfunctions vanish in a pair of Stokes sectors of opening angle $\frac{\pi}{2}$ centered about the positive-real and negative-real axes. These sectors are invariant under both $\cP$ and $\cPT$ reflections. However, as $\varepsilon$ increases, both sectors rotate {\it downward}; the right sector rotates in the negative direction and the left sector rotates 
in the positive direction until the upper edge of each sector lies on the real axis. The opening angles of each sector shrink from $\frac{\pi}{2}$ to $\frac{\pi}{3}$ as $\varepsilon$ increases from 0 to 2. The orientation of these sectors is no longer $\cP$ symmetric but remains $\cPT$ (left-right) symmetric.

\begin{figure}[t]
\center
\includegraphics[scale = 0.25]{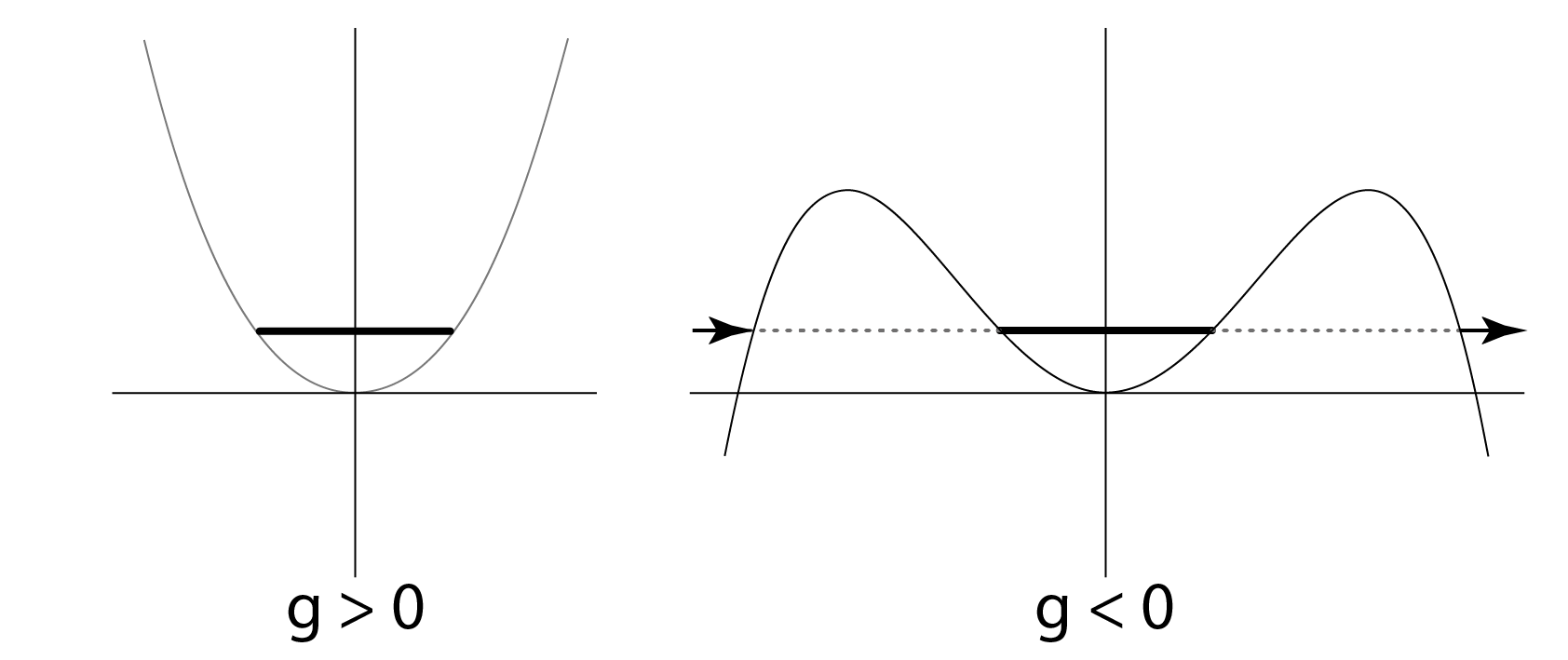}
\caption{Schematic diagram showing how a stable bound state in the right-side-up potential ${\hat x}^2+{\hat x}^2(i{\hat x})^\varepsilon$ at $\epsilon=0$ (left panel) remains stable as $\epsilon$ increases from $0$ to $2$. Even though the potential is upside-down at $\varepsilon=2$ (right panel) waves propagate out to $x=\pm\infty$ in a parity-time-symmetric (unidirectional) fashion (right panel); that is, incoming waves from the left balance out-going waves to the right, so the amplitude of the bound state in the potential neither grows nor decays in time.}
\label{Fig4}
\end{figure}

The schematic diagram in Fig.~\ref{Fig4} shows that a real energy level in the right-side-up potential of ${\hat H}_\varepsilon$ at $\varepsilon=0$ {\it continues to be real} when $\varepsilon=2$ even though the potential is upside down. This is because the boundary conditions at $x=\pm\infty$ are unidirectional \cite{pt189,pt595}. while the probability current flows out of the well on one side, there is an equal flow into the well on the other side, so the state in the potential well does not decay.

There is another way to perform a $\cPT$ deformation of the harmonic oscillator that {\it preserves the reality of the eigenvalues}. Instead of deforming the Hamiltonian (\ref{e8}), we can deform the $\cPT$-symmetric Hamiltonian
\begin{equation}
{\hat H}_\epsilon={\hat p}^2+{\hat x}^2+{\hat x}^2(-i{\hat x})^\varepsilon
\label{e9}
\end{equation}
by increasing $\varepsilon$ smoothly from $0$ to $2$. This deformation gives the same discrete positive-energy spectrum as in the $\varepsilon$ defomation above, but the wave configurations are opposite to those shown in Fig.~\ref{Fig4}; there are incoming waves at $x=+\infty$ and outgoing waves at $x=-\infty$. Each of these {\it unidirectional} wave configurations is separately invariant under $\cPT$ reflection. The energy levels and wave configurations can be observed in laboratory experiments~\cite{pt595}.

The $\cPT$-symmetric real-eigenvalue phase and the $\cP$-symmetric complex-eigenvalue phase of the Hamiltonian (\ref{e5}) are distinct physically observable phases of the Hamiltonian {\ref{e5}). The positive-coupling Hamiltonian (\ref{e7}) has a positive-spectrum phase that is both $\cP$- and $\cPT$-symmetric and the negative-coupling Hamiltonian (\ref{e5}) has a
$\cPT$-symmetric phase with a positive spectrum, but these two spectra are {\it different}. One spectrum cannot be obtained from the other by analytic continuation in the coupling constant $g$.

Furthermore, the Green's functions $G_n$ of the two Hamiltonians ${\hat H}={\hat p}^2+{\hat x}^2\pm g{\hat x}^4$ are completely different. The odd-$n$ Green's functions of the positive-$g$ theory vanish, but the odd-$n$ Green's functions are nonzero for the negative-$g$ $\cPT$-symmetric theory \cite{pt579}.

The energy levels of the simpler Hamltonian
\begin{equation}
{\hat H}={\hat p}^2+{\hat x}^2(i{\hat x})^\varepsilon  
\label{e10}
\end{equation}
are plotted in Fig.~\ref{Fig5} for $-1<\varepsilon\leq 4$. This is the earliest class of $\cPT$-symmetric Hamiltonians that was studied in detail \cite{pt1}. For $\varepsilon\geq0$ the eigenvalues of $\hat H$ in (\ref{e10}) are all real, positive, and discrete. This property was first discovered in \cite{pt1} and spectral reality was proved in \cite{pt274}.

\begin{figure}[t]
\center
\includegraphics[scale = 0.23]{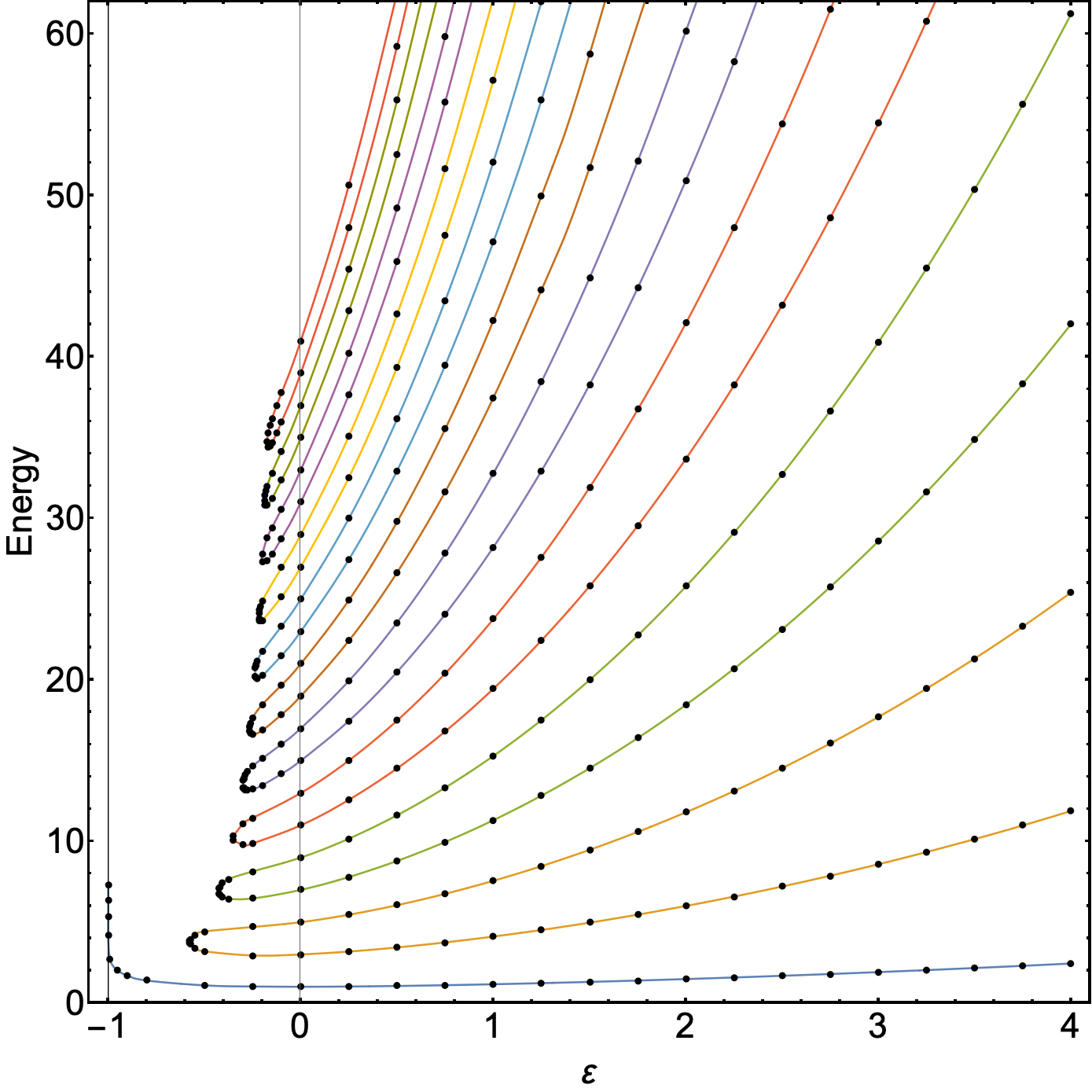}
\caption{First twenty-one eigenvalues of the $\cPT$-symmetric Hamiltonian $H$ in (\ref{e10}) plotted for $-1<\varepsilon\leq4$. The eigenvalues are all real, positive, discrete, and monotonically increasing for $\varepsilon\geq0$. Note that as $\varepsilon$ decreases below 0 some eigenvalues become degenerate pairwise and enter the complex plane as complex-conjugate pairs, but the ground-state energy continues to remain real and becomes singular as $\varepsilon$ approaches $-1$. (This plot only shows the eigenvalues when they are real.)}
\label{Fig5}
\end{figure}

Figure \ref{Fig5} indicates that there is a spectral transition at $\varepsilon=0$: When $\varepsilon<0$, some eigenvalues remain real but others occur in complex-conjugate pairs. (These complex eigenvalues are not shown in the figure.) This transition, which is called the $\cPT$ {\it transition}, has been observed in many laboratory experiments involving different kinds of $\cPT$-symmetric physical systems (see refs.~in \cite{pt717}).

\section{Quantum Hamiltonians with multiple phases} The number of phases defined by a quantum-mechanical Hamiltonian $\hat H$ equals the number of pairs of Stokes sectors in which one can pose an eigenvalue problem for the Schr\"odinger equation associated with $H$. For example, the cubic-potential Hamiltonian
\begin{equation}
{\hat H}={\hat p}^2+i{\hat x}^3.
\label{e11}
\end{equation}
has five Stokes sectors in the complex-$x$ plane in which the solutions to the Schr\"odinger equation
\begin{equation}
-\psi''(x)+ix^3\psi(x)=E\psi(x)
\label{e12}
\end{equation}
associated with $\hat H$ can vanish exponentially as $|x|\to\infty$. These sectors are labeled in Fig.~\ref{Fig6} as A, B, C, D, E.

\begin{figure}[t]
\center
\includegraphics[scale = 0.55]{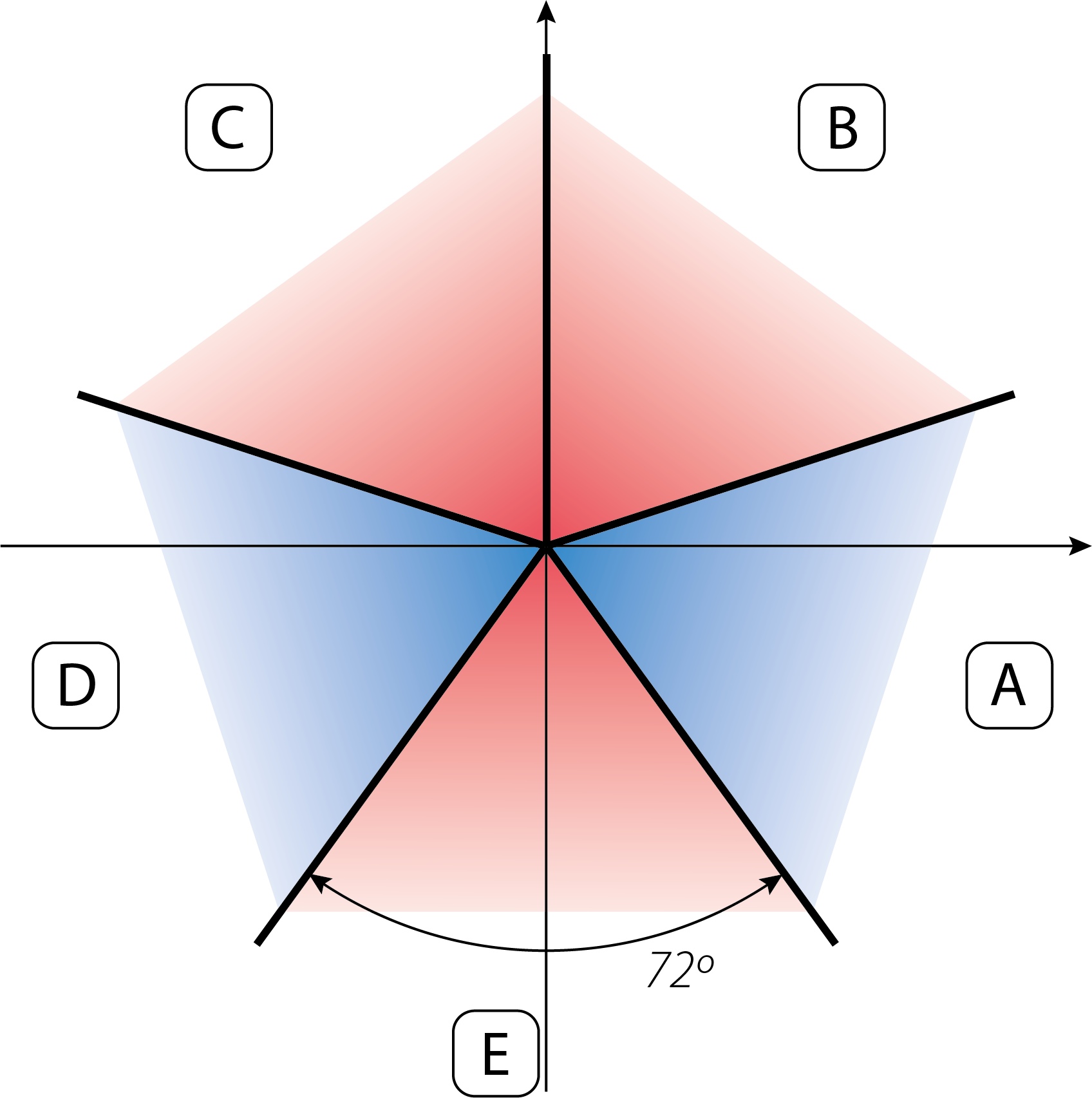}
\caption{Stokes sectors in the complex-$x$ plane in which solutions to the Schr\"odinger equation (\ref{e12}) may vanish exponentially as $|x|\to\infty$. Each sector has an angular opening of $72^\circ$.}
\label{Fig6}
\end{figure}
 
The eigenfunctions of $H$ in (\ref{e11}) must vanish exponentially in a {\it pair of noncontiguous Stokes sectors}. There are five such pairs of sectors: AC, AD, BD, BE, and CE. Thus, the Hamiltonian $\hat H$ defines five different phases. Each of these phases has an countably infinite set of eigenfunctions and corresponding eigenvalues. However, the eigenvalue problem associated with the AD pair of sectors is special because the eigenvalues are all {\it real and positive}; this is because the AD pair of Stokes sectors is $\cPT$ (left-right) symmetric. These real eigenvalues are shown in Fig.~\ref{Fig5} for the value $\varepsilon=1$. The eigenvalues associated the other four phases are all complex. 

Next, consider the sextic Hamiltonian
\begin{equation}
{\hat H}={\hat p}^2+{\hat x}^6.
\label{e13}
\end{equation}
There are eight Stokes sectors in the complex-$x$ plane in which the solutions to the Schr\"odinger equation
\begin{equation}
-\psi''(x)+x^6\psi(x)=E\psi(x)
\label{e14}
\end{equation}
associated with $\hat H$ can vanish like $\exp\left(-\frac{1}{4}|x|^4\right)$ as $|x|\to\infty$. The angular opening of each sector is $45^\circ$. The sectors are shown in Fig.~\ref{Fig7} and are labeled A, B, $\dots$, H.

\begin{figure}[t]
\center
\includegraphics[scale = 0.5]{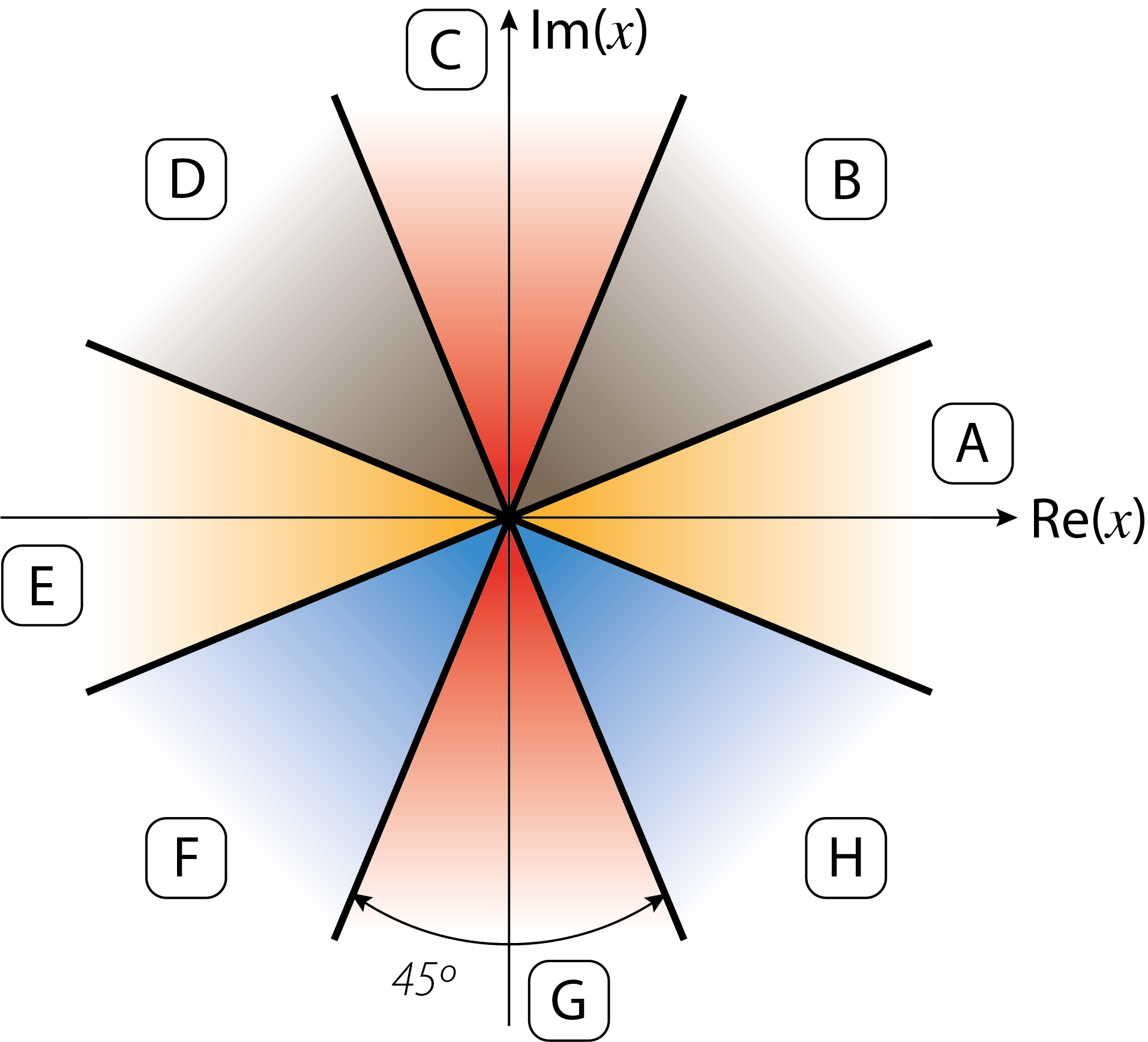}
\caption{Stokes sectors for the eigenfunctions of the Hamiltonian (\ref{e13}). If the eigenfunctions vanish exponentially in the AE pair of Stokes sectors, which contain the positive-real and negative-real axes, the associated eigenvalues are those of the conventional Hermitian $x^6$ potential. However, if the eigenfunctions vanish exponentially in the BD or FH pair of Stokes sectors, which are adjacent to and above or below the AE sectors, both sets of associated eigenvalues are those of the $\cPT$-symmetric sextic theory in (\ref{e10}) with $\varepsilon=4$. These eigenvalues are plotted in Fig.~\ref{Fig5}. The Hamiltonian (\ref{e13}) is real, so the Schr\"odinger eigenvalue problem is symmetric under complex conjugation and the array of Stokes sectors is up-down symmetric. Hence, the eigenvalues associated with the BD and the FH pairs of Stokes sectors are identical. Only these three $\cPT$-symmetric pairs of sectors correspond to real spectra. The three up-down symmetric DF, CG, and BH pairs of Stokes sectors give negative-real eigenvalues. The remaining 14 nonadjacent pairs of Stokes sectors give complex eigenvalues.} 
\label{Fig7}
\end{figure}

The lowest five eigenvalues in the Hermitan AE sectors are approximately $1.145$, $4.339$, $9.073$, $14.935$, and $21.714$, and the lowest five eigenvalues in the $\cPT$-symmetric BD and FH sectors are $2.439$, $11.882$, $25.412$, $42.024$, and $61.222$. WKB analysis reveals that for high energies the $\cPT$-symmetric eigenvalues are larger than the Hermitian eigenvalues by a factor of $2^{3/2}$. This feature might be analogous to families of particles having similar physical properties, such as the electron, muon, and tau, that have increasingly larger masses \cite{pt369}.

\section{Relationship between classical and quantum phases} The three positive-energy quantum phases of $\hat H$ in (\ref{e13}) have a clear classical analog. Figure \ref{Fig9} gives a plot of the classsical paths corresponding to $E=1$. Corresponding to the three left-right-symmetric pairs of quantum stokes sectors, there are three left-right-symmetric phases of classical paths. These phases fill the entire complex-$x$ plane and are bounded by separatrix paths.

\begin{figure}[t]
\center
\includegraphics[scale = 0.22]{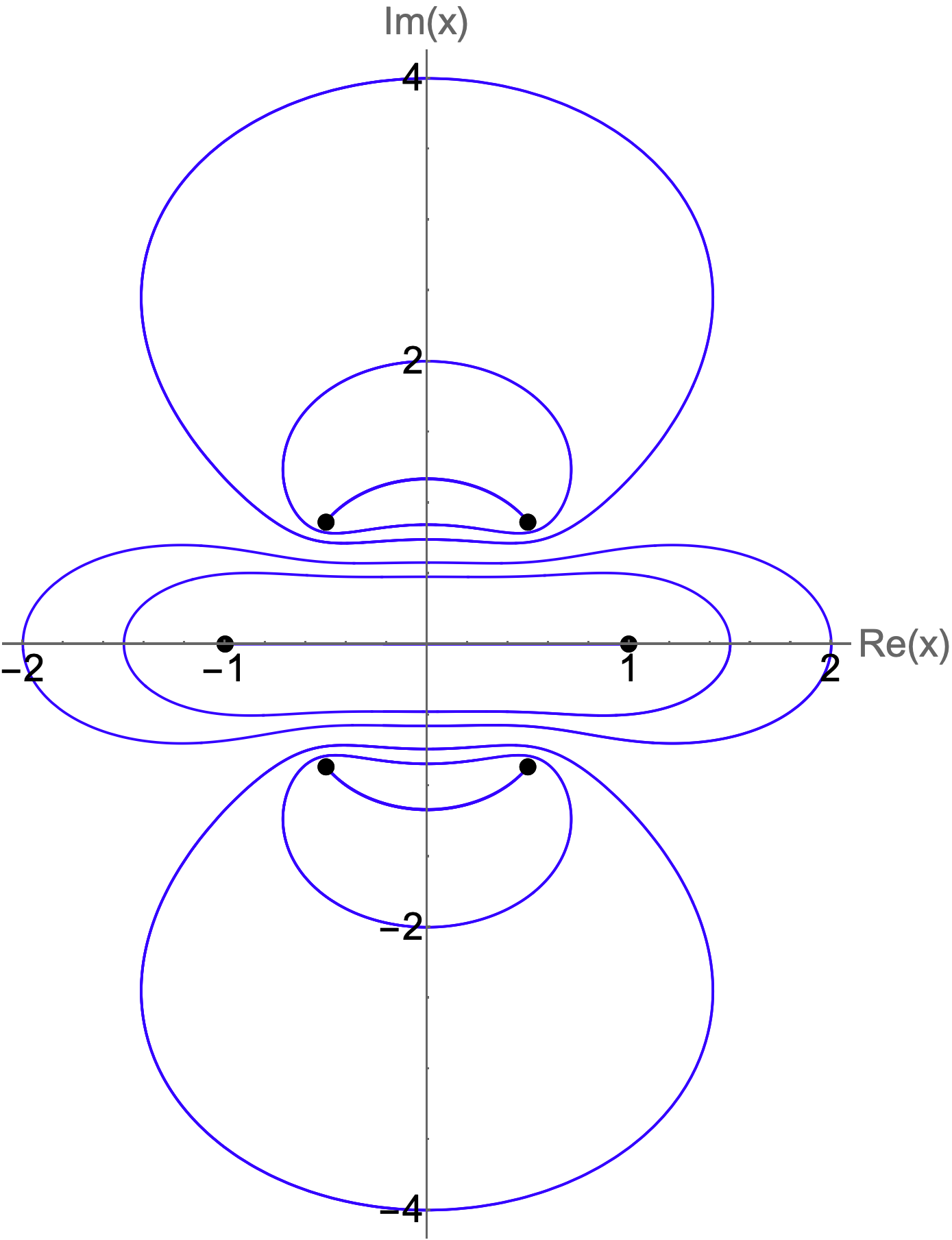}
\caption{Classical paths in the complex-$x$ plane for the sextic Hamiltonian $H=p^2+x^6$ for which the energy $E=1$. These paths are all closed and periodic. The classical turning points are indicated by black dots. The paths in the central phase all have the same period and are associated with the Hermitian spectrum of the quantum version of this Hamiltonian in (\ref{e13}). The upper and lower phases of classical paths all have the same period but this period is shorter than the period of the paths in the central phase because the quantum energies in the $\cPT$-symmetric phases are higher.}
\label{Fig8}
\end{figure}

There are also three phases of classical paths
having $E=-1$ corresponding to negative quantum energies. These phases are oriented vertically, corresponding to the vertical pairs of Stokes sectors BH, CG, and DF in Fig.~\ref{Fig7} and are bounded by separatrix paths.

\section{Phase transitions} For want of a better term, we use the term {\it phases} to describe the distinct sets of eigenfunctions and corresponding eigenvalues of a Hamiltonian. It is not yet clear if there is an order parameter that can be used to characterize these phases and if there are phase transitions. However, at the classical level one can arrange for particles in one phase to visit another phase. If the classical energy is complex, particle trajectories are no longer closed. Trajectories unwind and enter other phases, as shown in Fig.~\ref{Fig9}. (The average time for a classical particle to enter and leave a phase is proportional to $1/{\rm Im}\,E$, which is a classical analog of the time-energy uncertainty principle.) At the quantum level there are relationships between corresponding energy levels in different phases. For example, for $\hat H$ in (\ref{e11}), if $E_n^N$ is the $n$th energy level in the $N$th phase, then $E_n^1+E_n^2+E_n^3+E_n^4+E_n^5=0$.

\begin{figure}[t]
\center
\includegraphics[scale = 0.25]{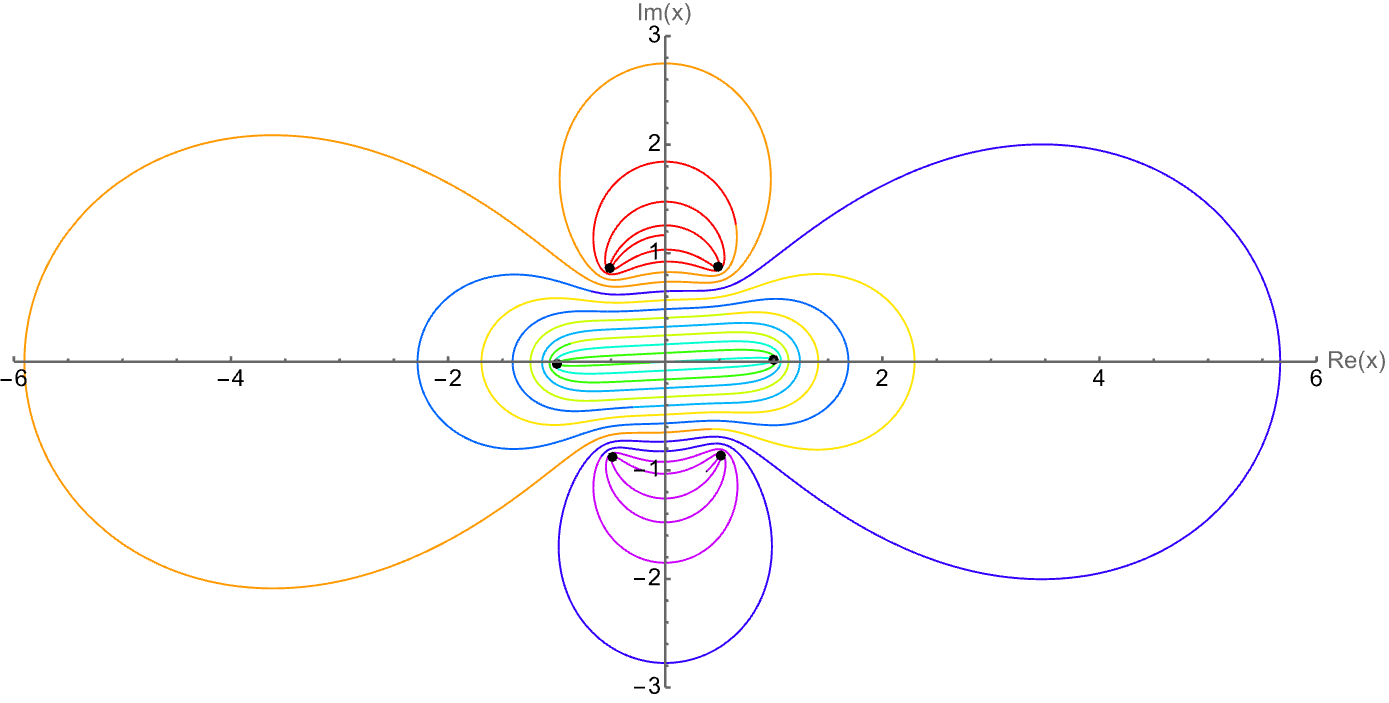}
\caption{Complex-energy version of Fig.~\ref{Fig8}. A classical particle having complex energy is not confined to one of the three regions of Fig.~\ref{Fig8}. Here $E=1+0.2\,i$, which breaks $\cPT$ symmetry and the pairs of turning points are no longer left-right symmetric. The particle is initially at $x(0)=1.167\,i$. This particle exhibits deterministic random behavior as it tunnels from region to region. The trajectory is uniquely determined by its initial position and the path of the particle never crosses itself. when the particle enters a region, it spirals inward and then outward, and then randomly enters an adjacent region. The trajectory of the particle in this figure is plotted from $t=0$ to $t=31.42$; the particle goes from the upper to the middle to the lower and back to the middle phase.}
\label{Fig9}
\end{figure}

\section{$\cPT$-symmetric quantum theory} In a phase in which the spectrum is entirely real and positive one can formulate a fully consistent quantum-mechanical theory that is based on a $\cPT$-symmetric Hamiltonian instead of a Hermitian Hamiltonian \cite{pt579,pt1,pt717}. Specifically, one can construct a Hilbert space of states with an appropriate positive-definite inner product. For example, there is a phase in which all of the energy levels of the deformed Hamiltonians in (\ref{e10}) are real and positive if $\varepsilon>0$ (see Fig.~\ref{Fig5}). Thus, despite its non-Hermiticity, ${\hat H}_\varepsilon$ defines a physically acceptable quantum-mechanical system.

\section{$\cPT$-symmetric quantum field theory}
This multiple-phase phenomenon extends from quantum mechanics to quantum field theory. For example, a $g\phi^4$ $(g<0)$ quantum field theory appears to have two phases, an unstable $\cP$-symmetric phase for scalar $\phi$ and a stable $\cPT$-symmetric phase for pseudoscalar $\phi$. The $g>0$ theory is not asymptotically free, but the $\cPT$-symmetric $g<0$ theory {\it is} asymptotically free \cite{pt586,pt711}.

Another example of a $\cPT$-symmetric quantum field theory is lurking in the classic 1952 paper on the divergence of perturbation series in quantum electrodynamics (QED) \cite{pt584}). This paper argued heuristically that a perturbation series in powers of the fine-structure constant $\alpha$ must diverge because if $\alpha$ were replaced by $-\alpha$, electrons and positrons would repel and the vacuum state would become unstable. Thus, the nature of the physics would change discontinuously at $\alpha=0$. The conclusion of this argument is correct, as shown in subsequent research \cite{pt588,pt590,pt589,pt598,pt641}, but the mathematical reason for this divergence is simply that the number of Feynman diagrams grows factorially with the order of perturbation theory.

There is an important subtlety not considered in Ref.~\cite{pt584}. Like the anharmonic-oscillator Hamiltonian (\ref{e5}), the negative sign of $\alpha$ in QED leads to {\it two} phases, (1) a $\cP$-symmetric phase (PQED) with an unstable vacuum state, and (2) a non-Hermitian $\cPT$-symmetric phase (PTQED) in which $A^\mu$ is an axial-vector field \cite{pt454}. Unlike conventional Hermitian QED, the PTQED phase is asymptotically free and appears to have a stable vacuum state.

The program of Johnson, Baker, and Willey for calculating $\alpha$ in {\it massless} QED failed because QED is not asymptotically free, as shown by the absence of a convergent sequence of positive zeros in the weak-coupling expansion of $F_1(\alpha)$ (the logarithmic-divergent part of the Beta function) \cite{pt713,pt715,pt716}:
$$F_1(\alpha)=\frac{4}{3}\left(\frac{\alpha} {4\pi}\right)+4\left(\frac{\alpha}{4\pi}\right)^2-2\left(\frac{\alpha}{4\pi}\right)^3-46\left(\frac{\alpha}{4\pi}\right)^4 +\cdots.$$
Even if this series is converted to Pad\'e approximants \cite{pt714,pt715,pt716, pt712,pt709,pt710}, no sequence of positive zeros emerges. However, for massless PTQED $F_1(\alpha)$ does appear to have a convergent sequence of {\it negative} zeros $\alpha=-4.187,\,-3.657,\,-3.590,\,\cdots$, which suggests that PTQED has an asymptotically free phase that is fundamentally different from Hermitian QED \cite{pt453}. 

Studies of the Casimir force support the possibility that QED has a PTQED phase. In QED a conducting spherical shell experiences a {\it repulsive} Casimir force that tends to inflate the sphere \cite{pt718}. However, in PTQED a conducting spherical shell tends to collapse. If this attractive force is balanced by placing an electric charge on the shell, one obtains the approximate {\it negative} $\alpha=-0.09235$ \cite{pt453}.

\section{Concluding remarks}Even though these field-theory results are not rigorous, it is now clear that studies of quantum theory in the complex domain will continue to reveal a rich array of new, unexpected, and experimentally observable behaviors. 

\acknowledgments
CMB thanks the Alexander von Humboldt and Simons Foundations for financial support. 

\bibliography{PT}

\begin{thebibliography}{10}
\expandafter\ifx\csname url\endcsname\relax\def\url#1{\texttt{#1}}\fi

\bibitem{pt486}
\Name{Bender C.~M. \and Hook D.~W.} \REVIEW{Journal of Physics A: Mathematical and Theoretical}{41}{2008}{24405}.

\bibitem{pt579}
\Name{Bender C.~M., Dorey P.~E., Dunning C., Fring A., Hook D.~W., Jones H.~F., Kuzhel S., L\'evai G. \and Tateo R.} \Book{PT Symmetry in Quantum and Classical Physics} (World Scientific, Singapore) 2019.

\bibitem{pt189}
\Name{Ramezani H., Kottos T., El-Ganainy R. \and Christodoulides D.~N.} \REVIEW{Physical Review A}{82}{2010}{043803}.

\bibitem{pt14}
\Name{Bender C.~M. \and Wu T.~T.} \REVIEW{Physical Review}{184}{1969}{1231}.

\bibitem{pt595}
\Name{Soley M.~B., Bender C.~M. \and Stone A.~D.} \REVIEW{Physical Review Letters}{130}{2023}{250404}.

\bibitem{pt1}
\Name{Bender C.~M. \and Boettcher S.} \REVIEW{Physical Review Letters}{80}{1998}{5243}.

\bibitem{pt274}
\Name{Dorey P., Dunning C. \and Tateo R.} \REVIEW{Journal of Physics A: Mathematical and General}{34}{2001}{5679}.

\bibitem{pt717}
\Name{Bender C.~M. \and Hook D.~W.} \REVIEW{ArXiv}{}{2023}{2312.17386}.

\bibitem{pt369}
\Name{Bender C.~M. \and Klevansky S.~P.} \REVIEW{Physical Review Letters}{105}{2010}{031601}.

\bibitem{pt586}
\Name{Symanzik K.} \REVIEW{Communications in Mathematical Physics}{45}{1975}{79}.

\bibitem{pt711}
\Name{Bender C.~M., Duncan A. \and Jones H.~F.} \REVIEW{Physical Review D}{49}{1994}{4219}.

\bibitem{pt584}
\Name{Dyson F.~J.} \REVIEW{Physical Review}{85}{1952}{631}.

\bibitem{pt588}
\Name{Hurst C.~A.} \REVIEW{Mathematical Proceedings of the Cambridge Philosophical Society}{48}{1952}{625}.

\bibitem{pt590}
\Name{Petermann A.} \REVIEW{Physical Review}{89}{1953}{1160}.

\bibitem{pt589}
\Name{Thirring W.} \REVIEW{Helvetica Physica Acta}{26}{1953}{33}.

\bibitem{pt598}
\Name{Glimm J. \and Jaffe A.} \REVIEW{Physical Review}{176}{1968}{1945}.

\bibitem{pt641}
\Name{Bender C.~M. \and Wu T.~T.} \REVIEW{Physical Review Letters}{27}{1971}{461}.

\bibitem{pt454}
\Name{Bender C.~M., Cavero-Palaez I., Milton K.~A. \and Shajesh K.~V.} \REVIEW{Physics Letters B}{613}{2005}{97}.

\bibitem{pt713}
\Name{Jost R. \and Luttinger J.~M.} \REVIEW{Helvetica Physica Acta}{23}{1950}{201}.

\bibitem{pt715}
\Name{Rosner J.~L.} \REVIEW{Annals of Physics}{44}{1967}{11}.

\bibitem{pt716}
\Name{Gorishny S.~G., Kataev A.~L., Larin S.~A. \and Surguladze L.~R.} \REVIEW{Physics Letters B}{256}{1991}{81}.

\bibitem{pt714}
\Name{Rosner J.} \REVIEW{Physical Review Letters}{17}{1966}{1190}.

\bibitem{pt712}
\Name{Johnson K., Baker M. \and Willey R.~S.} \REVIEW{Physical Review Letters}{11}{1963}{518}.

\bibitem{pt709}
\Name{Johnson K., Baker M. \and Willey R.} \REVIEW{Physical Review}{136}{1964}{B1111}.

\bibitem{pt710}
\Name{Johnson K., Baker M. \and Willey R.} \REVIEW{Physical Review}{163}{1967}{1699}.

\bibitem{pt453}
\Name{Bender C.~M. \and Milton K.~A.} \REVIEW{Journal of Physics A: Mathematical and General}{32}{1999}{L87}.

\bibitem{pt718}
\Name{Boyer T.~H.} \REVIEW{Physical Review}{174}{1968}{1764}.

\end{thebibliography}
\end{document}